\documentclass[twocolumn,aps,prl,superscriptaddress,showpacs,showkeys]{revtex4}

\usepackage{CJK}%
\usepackage{color}%
\usepackage{amsmath,bm}%
\usepackage{graphicx}%
\begin{document}
\begin{CJK*}{GBK}{}

\title{Sensitivity study of experimental measures for nuclear liquid-gas phase transition in statistical multifragmentation model (SMM)}
\author{W. Lin}
\affiliation{Key Laboratory of Radiation Physics and Technology of the Ministry of Education, Institute of Nuclear Science and Technology, Sichuan University, Chengdu 610064,	China}
\affiliation{Institute of Modern Physics, Chinese Academy of Sciences, Lanzhou, 730000, China}
\author{P. Ren}
\affiliation{Key Laboratory of Radiation Physics and Technology of the Ministry of Education, Institute of Nuclear Science and Technology, Sichuan University, Chengdu 610064,	China}
\affiliation{Institute of Modern Physics, Chinese Academy of Sciences, Lanzhou, 730000, China}
\author{H. Zheng}
\email[E-mail at:]{zheng@lns.infn.it}
\affiliation{Laboratori Nazionali del Sud, INFN, I-95123 Catania, Italy}
\author{X. Liu}
\email[E-mail at:]{liuxingquan@impcas.ac.cn}
\affiliation{Institute of Modern Physics, Chinese Academy of Sciences, Lanzhou, 730000, China}
\author{M. Huang}
\affiliation{College of Physics and Electronics information, Inner Mongolia University for Nationalities, Tongliao, 028000, China}
\author{R. Wada}
\affiliation{Cyclotron Institute, Texas A$\&$M University, College Station, Texas 77843}
\author{G. Qu}
\affiliation{Key Laboratory of Radiation Physics and Technology of the Ministry of Education, Institute of Nuclear Science and Technology, Sichuan University, Chengdu 610064,	China}
\affiliation{Institute of Modern Physics, Chinese Academy of Sciences, Lanzhou, 730000, China}
\date{\today}

\begin{abstract}
  The experimental measures of the multiplicity derivatives, the moment parameters, the bimodal parameter, the fluctuation of maximum fragment charge number (NVZ), the Fisher exponent ($\tau$) and Zipf's law parameter ($\xi$), are examined to search for the liquid-gas phase transition in nuclear multifragmention processes within the framework of the statistical multifragmentation model (SMM). The sensitivities of these measures are studied. All these measures predict a critical signature at or near to the critical point both for the primary and secondary fragments. Among these measures, the total multiplicity derivative and the NVZ provide accurate measures for the critical point from the final cold fragments as well as the primary fragments. The present study will provide a guide for future experiments and analyses in the study of nuclear liquid-gas phase transition.
\end{abstract}
\pacs{25.70-z, 24.10Cn}

\keywords{nuclear liquid-gas phase transition, experimental measures, statistical multifragmentation model (SMM)}

\maketitle
\end{CJK*}

\section*{I. Introduction}
 
 The interest in the multifragmentation processes, which was predicted long time ago~\cite{Bohr1936Nature} and has been extensively studied following the advent of 4$\pi$ detectors~\cite{Borderie2008PPNP,Gulminelli2006EPJA,Chomaz2004PR}, lies in the fact that it provides a wealth of information on nuclear dynamics, on the properties of the nuclear equation of state (EOS), and on the possible nuclear liquid-gas phase transition. The nuclear liquid-gas phase transition in multifragmentation process was first suggested in the early 1980s~\cite{Finn1982PRL,Minich1982PLB,Hirsch1984NPA}. It is  expected to occur when the nucleus is heated to a moderate temperature and breaks up on a short time scale into light particles and intermediate mass fragments with Z $\geq$ 3 (IMF). 
 
 In the past three decades, many experimental and theoretical works have been devoted to searching for the liquid-gas phase transition in the Fermi energy heavy-ion collisions and relativistic energy projectile fragmentations. Among the measures used for studies are the nuclear specific heat capacity (the caloric curves)~\cite{Suraud1989PPNP,Gross1993PPNP,Hagel1988NPA,Wada1989PRC,Cussol1993NPA,Pochodzalla1995PRL,Wada1997PRC,Hagel2000PRC,Furuta2006PRC}, the bimodality in charge  asymmetry~\cite{Lopez2005PRL,Pichon2006NPA,Borderie2010NPA}, the Fisher droplet model analysis~\cite{Fisher1967RPP,Elliott2002PRL,Ma2005PRC,Ma2005NPA,Ma2004PRC,Huang2010PRC,Giuliani2014PPNP}, the Landau free energy approach~\cite{Bonasera2008PRL,Huang2010PRC,Giuliani2014PPNP,Tripathi2011PRC,Tripathi2011JPCS,Tripathi2012IJMPE,Mabiala2013PRC}, the moment of the charge distributions~\cite{Campi1988PLB,Campi1986JPA,Das1998PRC,Mastinu1998PRC,Ma2005PRC}, the fluctuation properties of the heaviest fragment size (charge)~\cite{Mastinu1998PRC,Ma2005PRC,Ma2005NPA,Botet2001PRL,Frankland2005PRC}, the Zipf's law~\cite{Ma1999PRL,Ma1999EPJA}, the multiplicity derivatives recently proposed by S. Mallik et al.~\cite{Mallik2017PRCR} and the derivative of cluster size~\cite{Das2018Arxiv}. With these features, many considerable progresses have been accomplished on the theoretical as well as on the experimental side for the nuclear liquid-gas phase transition. Y.G. Ma et al. in Refs.~\cite{Ma2005PRC,Ma2005NPA,Ma2004PRC}, most of these measures except the multiplicity derivatives, were examined as a function of the excitation energy, using  rather light reaction systems of $^{40}$Ar+$^{27}$Al,$^{48}$Ti and $^{58}$Ni at 47 MeV/nucleon, and showed that all of them show a critical behavior around $E^{*}/A \sim  5.6$ MeV. However since all values of the measures are plotted as a function of the excitation energy, the signature appears as a broad peak around  $E^{*}/A \sim  5.6$ MeV. Therefore, the specific properties of the nuclear liquid-gas phase transition in hot nuclear matter are still under debate and many efforts are still required.
 
 In order to search for suitable observables in heavy-ion collisions, which can provide strong signatures for the nuclear liquid-gas phase transition and be a guide for future experiments, we investigate several experimental measures including the multiplicity derivatives, the moment parameters, the Zipf's law, and analyze the sensitivity of each observable in the framework of the statistical multifragmentation model (SMM)~\cite{ZhangNPA1987I,ZhangNPA1987II,Bondorf95,Botvina2001,Soulioutis2007,Lin2018PRC}. SMM is rather successful in describing the multiple production of intermediate mass fragments~\cite{Botvina1995NPA,DAgostino1996PLB,DAgostino1999NPA} and exhibits a phase transition of the liquid-gas type~\cite{Buyukcizmeci2005EPJA,Ogul2005NPA}.
 This article is organized as follows: A brief description of SMM is presented in Sec. II. The SMM calculations and analyses of phase transition are given in Sec. III. The discussions are given in Sec. IV. A brief summary is given in Sec. V.

\section*{II. Statistical multifragmentation model (SMM)}

 In SMM, the fragmenting system is in the thermal and chemical equilibrium at low density~\cite{Bondorf95,Botvina2001,Soulioutis2007,Lin2018PRC}. A Markov chain is generated to represent the whole partition ensemble in the version discussed below~\cite{Botvina2001}. All breakup channels (partitions) for nucleons and excited fragments are considered under the conservation of mass, charge, momentum and energy. The primary fragments are described by liquid-drops at a given freezeout volume. Light clusters with mass number A $\le$ 4 are considered as stable particles (``nuclear gas''). Their masses and spins are taken from the experimental values. Only translational degrees of freedom of these particles are taken into account in the entropy of the system. When the nuclear density becomes very low, the binding energy of clusters is significantly modified by the Pauli blocking and clusterization~\cite{Typel2010PRC}, but these effects are not taken into account in the SMM. Fragments with A $>$ 4 are treated as spherical excited nuclear liquid drops and the free energies $F_{A,Z}$ are given as a sum of the bulk, surface, Coulomb, and symmetry-energy contributions,
 \begin{eqnarray}\label{eq:FreeEnergy}
 F_{A,Z} &=& F^{B}_{A,Z}+F^{S}_{A,Z}+E^{C}_{A,Z}+F^{sym}_{A,Z},
 \end{eqnarray}
 where
 \begin{eqnarray}
 \label{eq:VolumeFreeE}
 F^{B}_{A,Z} &=& (-W_{0}-T^{2}/\varepsilon_{0})A,\\
 F^{S}_{A,Z} &=& B_{0}A^{2/3}\left[ \frac{T^{2}_{c}-T^{2}}{T^{2}_{c}+T^{2}} \right]^{5/4},\\
 E^{C}_{A,Z} &=& \frac{3}{5}\frac{e^2}{r_{0}}[1-(\rho/\rho_0)^{1/3}]\frac{Z^{2}}{A^{1/3}},\\
 F^{sym}_{A,Z} &=& \gamma (A-2Z)^{2}/A - TS^{sym}_{A,Z}.
 \label{eq:SymmetryEnergy}
 \end{eqnarray}
 $W_{0}$ = 16 MeV is used for the binding energy of infinite nuclear matter, and $\varepsilon_{0}$ = 16 MeV is related to the level density; $B_{0}$ = 18 MeV is used for the surface coefficient; $T_{c}$ = 18 MeV is used for the critical temperature of infinite nuclear matter; $e$ is the charge unit and $r_0$ = 1.17 fm; $\rho$ is the density at the breakup and $\rho_0$ is the normal nuclear density; $\gamma$ is the symmetry energy parameter; the $S^{sym}_{A,Z}$ is the symmetry entropy of fragment introduced in our previous work~\cite{Lin2018PRC}.
 
 The entropy of fragments $S_{A,Z}$ can be derived from the free energy as
 \begin{eqnarray}
 S_{A,Z} = -\frac{\partial F_{A,Z}}{\partial T} = S^{B}_{A,Z}+S^{S}_{A,Z}+S^{sym}_{A,Z}.
 \label{eq:S_AZ}
 \end{eqnarray}
  
 After the primary breakup, the Coulomb acceleration and the secondary de-excitation are performed to get the final secondary fragments. In the de-excitation processes, the Fermi break-up of light primary fragments ($A<16$), the successive particle emission ($A>16$) and the fission of heavy nuclei ($A>200$) are taken into account.
 
\section*{III. SMM calculations and analyses of phase transition}
 
 SMM calculations are performed with the source mass number $A_s$ = 100, charge number $Z_s$ = 45, the fragmenting volume $V = 6V_0$, where $V_0$ is the volume at the normal nuclear density. The default symmetry energy coefficient $\gamma$ = 25 MeV is used. The input source excitation energy ($E_x$) varies from 1 to 15 MeV/nucleon with the energy step of 0.25 MeV/nucleon. More than 1 Million events are generated for each $E_x$. In order to be a guide in future experiments, the calculations are performed both for the primary and secondary fragments. 
 
 In SMM the ``temperature'' depends slightly on the fragmenting channel because of the energy fluctuates from partition to partition with the Markov-chain method. The energies are determined from the energy balance for a given partition. Therefore, the average value over all exit channels is used as the source temperature in the following analyses~\cite{Lin2018PRC}.

 \begin{figure}[hbt]
	\centering
	\includegraphics[scale=0.44]{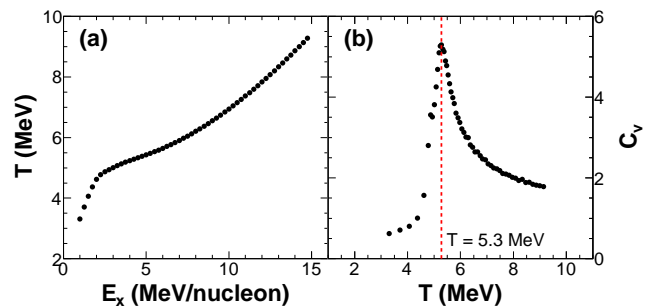}
	\caption{\footnotesize
		(Color online) (a) The caloric curve of fragmenting source with $A_s$ = 100, $Z_s$ = 45 of SMM calculations. (b) The specific heat capacity $C_{v}$ derived from caloric curve as a function of source temperature. The vertical line shows the critical point at $T = $ 5.3 MeV.
	}		
	\label{fig:fig01_caloric}
 \end{figure}

 The specific heat capacity has long been considered to be a measure that should provide important information on the postulated nuclear liquid-gas phase transition~\cite{Suraud1989PPNP,Gross1993PPNP,Natowitz2002PRC,Agostino2000PLB,Agostino2002NPA}. As one can see from Fig.~\ref{fig:fig01_caloric} (a) that a notable plateauing of the caloric curve is observed at $E_x \sim 4$ MeV for the SMM calculations, which results in a sharp increasing of the specific heat capacity, $C_v$, as shown in Fig.~\ref{fig:fig01_caloric} (b). The sharp maximum of the $C_v$ strongly suggests that the liquid-gas phase transition occurs in SMM. The critical point at temperature $T = $ 5.3 MeV is obtained. Experimentally the caloric curve has been measured in many experiments.  The plateau of caloric curve is qualitatively observed at the excitation energy of 5 - 10 MeV, depending on the system size~\cite{Natowitz2002PRC}. However, due to the complexity of reaction mechanisms and sequential secondary decay processes, the experimental determination of the excitation energy and temperature becomes inaccurate and do not allow to determine the critical point as a sharp transition even if it is there. Therefore, it is crucial to find a good thermometer for the experiments, which enables us to determine the temperature reliably and accurately and has minimal effect from the sequential decay process. Here we will use $T = $ 5.3 MeV as the theoretical critical point for the reference.

\subsection*{A. Multiplicity derivatives} 
 
 The derivatives of total multiplicity and IMF multiplicity were recently proposed as an observable to search for nuclear liquid-gas phase transition by S. Mallik et al. in Ref.~\cite{Mallik2017PRCR}. They showed that the multiplicity derivatives show a strong signature marked for the first-order phase transition in the canonical thermodynamic model (CTM)~\cite{Das2005PR}, which is claimed to be essentially same as SMM.
 
\begin{figure}[hbt]
	\centering
	\includegraphics[scale=0.44]{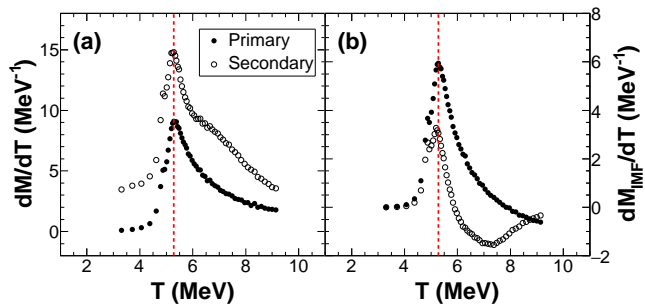}
	\caption{\footnotesize
		(Color online) (a) The total multiplicity derivatives of fragmenting system of SMM versus the source temperature. (b) Similar to that in (a) but for the derivatives of IMF multiplicity. The solid and open circles correspond to that of primary and secondary fragments, respectively. The vertical lines indicate the critical point at $T = $ 5.3 MeV from Fig.~\ref{fig:fig01_caloric} (b).
	}		
	\label{fig:fig02_dMdT}
 \end{figure}

 We apply the multiplicity derivatives to the fragmenting system calculated by SMM. Fig.~\ref{fig:fig02_dMdT} (a) and (b) show the total  and IMF multiplicity derivatives as a function of source temperature, respectively for both primary (solid circles) and the secondary (open circles) fragments. All distributions show a sharp increasing and have a maximum at or near the critical temperature of $T = $ 5.3 MeV shown by vertical lines in both figures. The good agreement between critical temperatures in the multiplicity derivatives and that in the specific heat capacity is found. The fact that only slight lower value ($\sim$ 0.1 MeV) is found in IMF multiplicity derivative of secondary fragments, indicates that the multiplicity derivatives provide a good measure in searching for the critical point of nuclear matter liquid-gas phase transition. Our results are consistent with the conclusions in Ref.~\cite{Mallik2017PRCR}.

\subsection*{B. Moment parameters}
 
 The general definition of the $k$th moment~\cite{Campi1988PLB,Campi1986JPA,Ma2005PRC} of charge distribution is given as
 \begin{equation}\label{eq:moment}
 M_k = \sum_{Z_i\neq Z_{max}}{n_iZ_i^k},
 \end{equation}
 where $n_i$ is the multiplicity of fragment with charge number $Z = Z_i$ in each event. Using the zeroth ($M_0$), first ($M_1$) and second ($M_2$) moments, the quantity $\gamma_2$ is defined as
 \begin{equation}\label{eq:gamma2}
 \gamma_2 = \frac{M_2M_0}{M^2_1}.
 \end{equation}
 
 \begin{figure}[hbt]
 	\centering
 	\includegraphics[scale=0.44]{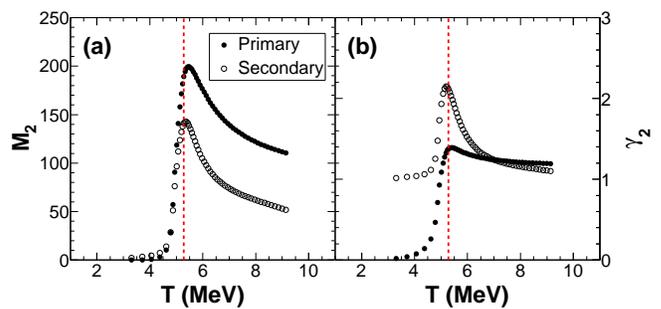}
 	\caption{\footnotesize
 		(Color online) (a) $M_2$ as a function of source temperature. (b) $\gamma_2$ as a function of source temperature. Solid and open circles correspond to that of primary and secondary fragments, respectively. The vertical lines indicate the critical point at $T = $ 5.3 MeV from Fig.~\ref{fig:fig01_caloric} (b).
 	}		
 	\label{fig:fig03_Moments}
 \end{figure}

 $M_2$ and $\gamma_2$ are expected to show the critical point at which the fluctuations in fragment sizes become the largest~\cite{Ma2005PRC,Campi1986JPA,Campi1988PLB}.  Fig.~\ref{fig:fig03_Moments} (a) and (b) show the results of $M_2$ and $\gamma_2$ as a function of source temperature, respectively. As one can see from Fig.~\ref{fig:fig03_Moments} (a), the $M_2$ of primary fragments shows a maximum at slightly higher ($\sim$ 0.2 MeV) temperature than the critical temperature $T = $ 5.3 MeV, whereas the maximum of $M_2$ for secondary fragments is the same as the critical temperature. The maximum value of $\gamma_2$ of primary fragments appears at the temperature slightly larger than the critical temperature. On the contrary, the maximum of $\gamma_2$ of secondary fragments is slightly lower than the critical temperature. The deviations of those in both the primary and secondary fragments are less than 0.1 MeV as shown in Fig.~\ref{fig:fig03_Moments} (b).

\subsection*{C. Bimodal parameter and Fluctuations of maximum fragments}

 The bimodality~\cite{Lopez2005PRL,Pichon2006NPA,Borderie2010NPA} is a double peaked distribution of an order parameter, which comes from the anomalous convexity of the underlying microcanonical entropy. It can be interpreted as the coexistence of different phases in the system and provides a definition of an order parameter as the best variable to separate the two maxima of the distribution~\cite{Borderie2002JPG}. In this framework when
 a nuclear system is in the coexistence region, the probability distribution of the order parameter becomes bimodal. In Ref.~\cite{Borderie2002JPG}, the sorting parameter with fragment atomic number $Z$ = 12 as a limit between two phases, $\left(\sum_{Z_i\geq13}Z_i-\sum_{3\leq Z_i\leq12}Z_i\right)/\sum_{Z_i\geq3}Z_i$, which may connect with the density difference of the two phases ($\rho_L$ - $\rho_G$), was chosen as the order parameter in the analysis of INDRA data. 
 
\begin{figure}[hbt]
	\centering
	\includegraphics[scale=0.44]{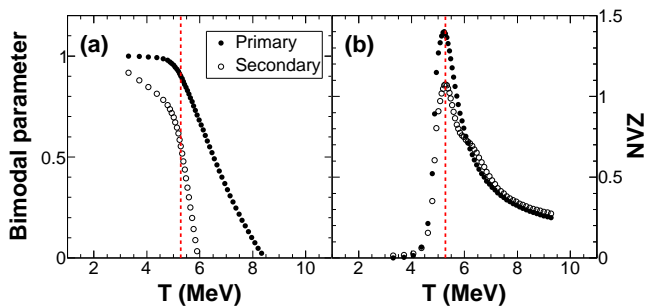}
	\caption{\footnotesize
		(Color online) (a) The bimodal parameter as a function of source temperature. (b) The NVZ as a function of source temperature. Solid and open circles correspond to that of primary and secondary fragments, respectively. The vertical lines indicate the critical point at $T = $ 5.3 MeV from Fig.~\ref{fig:fig01_caloric} (b).
	}		
	\label{fig:fig04_Bimodal}
 \end{figure}

 As pointed out by Ma et al. in Ref.~\cite{Ma2005PRC}, the $Z$ limit should be reduced between two phases for light systems and the critical temperature appears at the inflection point of the bimodal parameter. In the present analysis, we choose $Z = 3$ as the limit between the two phases, and therefore the bimodal parameter can be defined as $\left(\sum_{Z_i\geq4}Z_i-\sum_{1\leq Z_i\leq3}Z_i\right)/\sum_{Z_i\geq1}Z_i$. Fig.~\ref{fig:fig04_Bimodal} (a) shows the bimodal parameter as a function of source temperature. Lower temperatures of inflection point in bimodal parameter are found both for primary and secondary fragments compared to the critical temperature $T =$ 5.3 MeV. This behavior is generally observed for the bimodal parameter when the sorting limit $Z$ changed to $4 < Z < 12$ and/or the light charged particles ($Z \leq 2$) from the sorting fragments are excluded.
 
 The fluctuation of order parameter proposed by Botet in Ref.~\cite{Botet2000PRE}, provides a method to select an order parameter and characterize critical and off-critical behaviour, without any equilibrium assumption. The fluctuations in the atomic number of largest fragment ($Z_{max}$) have been applied in the analysis of INDRA data in Ref.~\cite{Borderie2002JPG} and the normalized variance of $Z_{max}$ (NVZ) was utilized by Dorso et al., in Ref.~\cite{Dorso1999PRC} to investigate the fluctuations of $Z_{max}$, which is given as
 \begin{equation}
 NVZ = \frac{\sigma^2_{Z_{max}}}{\left\langle Z_{max}\right\rangle}.
 \end{equation} 
 In the SMM calculations, the $Z_{max}$ does not always show a Gaussian distribution. Therefore, we apply the root-mean-square (RMS) of $Z_{max}$ as $\sigma_{Z_{max}}$ in NVZ. Fig.~\ref{fig:fig04_Bimodal} (b) shows the NVZ as a function of source temperature. One can see that the maxima of NVZ for both the primary and secondary fragments appear at the same temperature as that of the critical point, indicates that the NVZ also provides a good measure in searching for the critical point of nuclear matter liquid-gas phase transition.

\subsection*{D. Fisher exponent and Zipf's law parameter}

 The modified Fisher model (MFM)~\cite{Fisher1967RPP,Bonasera2000RNCSIF,Huang2010PRC,Lin2014PRCR} has been extensively applied to the analysis of multifragmentation events since it was first adopted by Purdue's group in Refs.~\cite{Minich1982PLB,Hirsch1984NPA,Hirsch1984PRC}. The fragment mass distributions in multifragmentation events are well described by a power law distribution of $A^{-\tau}$ with the power-law exponent $\tau \sim 2.3$~\cite{Fisher1967RPP,Bonasera2008PRL,Huang2010PRC}.
 
\begin{figure}[hbt]
	\centering
	\includegraphics[scale=0.44]{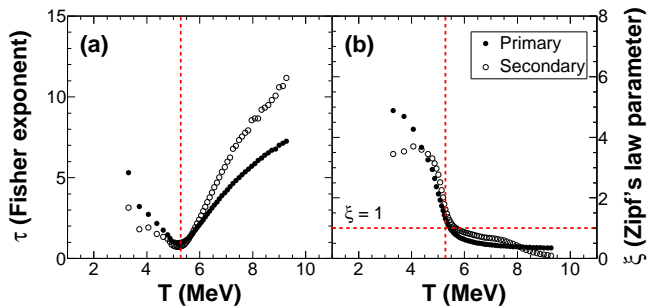}
	\caption{\footnotesize
		(Color online) (a) The Fisher exponent ($\tau$) extracted from Z distribution as a function of source temperature. (b) The Zipf's law parameter ($\xi$) as a function of source temperature. Solid and open circles correspond to that of primary and secondary fragments, respectively. The vertical lines indicate the critical point at $T = $ 5.3 MeV from Fig.~\ref{fig:fig01_caloric} (b). The horizontal line in (b) shows $\xi = 1$.
	}		
	\label{fig:fig05_Tau}
 \end{figure}

 In the framework of the MFM, the isotope yield in a multifragmentation reaction can be given as
 \begin{equation}\label{eq:MFM}
 Y(A,Z) = Y_0A^{-\tau}\exp\left[-\frac{F(A,Z)-\mu_nN-\mu_pZ}{T}\right],
 \end{equation}
 where $F(A,Z)$ is the free energy of fragment with mass A and charge Z, $\mu_n$ ($\mu_p$) is the neutron (proton) chemical potential. At the critical point, the exponential term in Eq.\eqref{eq:MFM} vanished and the distribution becomes a pure power law as
 \begin{equation}
 Y(A) = Y_0A^{-\tau}.
 \end{equation}
 As shown by Ogul in Ref.~\cite{Ogul2002PRCR}, the power law exponent of mass and charge distributions behave in a very similar fashion. Thus, we use the $Z^{-\tau}$ to fit the charge distribution. To avoid the contributions from the fission-like large fragments ($Z>20$) and from the coalescence-like small clusters ($Z \leq 2$), we adopt the same range of $Z = 5 - 15$ in the fit as that in Ref.~\cite{Botvina2006PRC}. The extracted power law exponents are shown in Fig.~\ref{fig:fig05_Tau} (a) both for the primary and secondary fragments. The minima of power law exponents appear at slightly lower ($\sim$ 0.15 MeV) temperature both for the primary and secondary fragments compared to the critical temperature $T = $ 5.3 MeV.

 The fragments hierarchy distribution gives another measure proposed by Ma in Refs.~\cite{Ma1999PRL,Ma1999EPJA}, which provides a method to search for the liquid-gas phase transition in a finite system. It can be defined by the so-called Zipf plot, which is a plot of the relationship between mean sizes of fragments rank-ordered in size (i.e., the largest fragment, the second large fragment, the third large fragment and so on). Originally the Zipf plot was used to analyze the hierarchy of usage of words in a language~\cite{Zipf1949}. It has been applied in a broad variety of areas, such as population distributions, the size distribution of cities, the distribution in strengths of earthquakes, etc.. The existence of very similar linear hierarchy distributions in these very different fields indicates that Zipf's law is a reflection of self-organized criticality~\cite{Turcotte1999RPP}.

 We apply the Zipf's law to the SMM events, in which the fragment charge number is employed as the variable to make a Zipf-type plot and the resultant distributions are fitted with a power law,
 \begin{equation}
 \left\langle Z_{rank}\right\rangle \propto rank^{-\xi}
 \end{equation}
 where $rank = i$ for the $i$th largest fragment. $\xi$ is the Zipf's law parameter. When $\xi \sim 1$, the Zipf's law is satisfied. The extracted $\xi$ values are plotted as a function of source temperature in Fig.~\ref{fig:fig05_Tau} (b) both for the primary and secondary fragments. One can see from the figure that the Zipf's law is satisfied at the temperature slightly larger than the critical temperature both for the primary and secondary fragments.

\section*{IV. Discussions}

 In section III, we investigate several experimental measures, which provide signatures for nuclear liquid-gas phase transition in heavy ion collisions in the framework of SMM both for the primary and secondary fragments. All these measures predict a critical temperature at or near to that from the specific heat capacity when they are plotted as a function of the source temperature. From the experimental point of view, it is important to provide measures which show same signature for the primary and secondary particles in the study of the nuclear liquid-gas phase transition.
 
 \begin{figure}[hbt]
 	\centering
 	\includegraphics[scale=0.35]{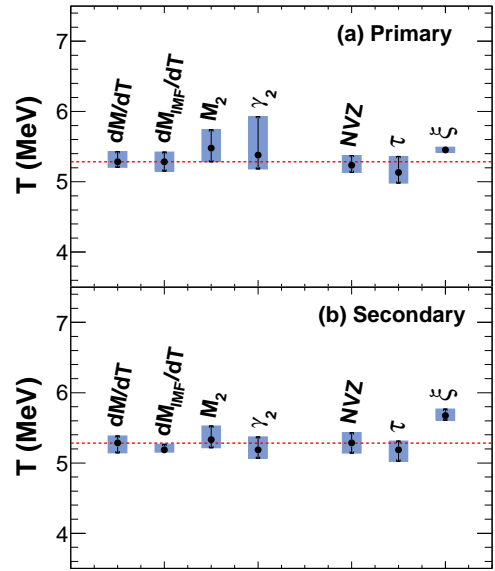}
 	\caption{\footnotesize
 		(Color online) (a) The sensitivities in the critical temperature of primary fragments for all the measures except for bimodal parameter. (b) The same as that in (a) but of secondary fragments. The horizontal lines indicate the critical point at $T = $ 5.3 MeV from Fig.~\ref{fig:fig01_caloric} (b). Solid circles correspond to the critical temperature extracted by each measure. For the error bars shown by the shaded area, see the detail in the text.
 	}		
 	\label{fig:fig06_Sensitivity}
 \end{figure}
 
 Due to the experimental errors (statistical and systematic), there would be some uncertainties included in the measures. Therefore, the sensitivity of these measures are further studied. The uncertainty, $\Delta T$, is evaluated as a quantitative measure when 5\% deviation from the maximum or minimum value is observed at $T = T_{c} \pm \Delta T$. Fig.~\ref{fig:fig06_Sensitivity} shows the sensitivities of all these measures except for the bimodal parameter, in which the inflection point is used to obtain the critical point and does not show the minimum or maximum value. One can easily get from the figure that the total multiplicity derivative and NVZ have the same critical temperature as that of the specific heat capacity both for the primary and secondary fragments. Moreover, the small errors in temperature in Fig.~\ref{fig:fig06_Sensitivity} (a) and (b) indicate that the total multiplicity derivative and NVZ provide the best measure in the study of nuclear liquid-gas phase transition.
 
 All the other measures are noticeably affected by the secondary decay as one can see in Fig.~\ref{fig:fig06_Sensitivity} (a) and (b). The critical temperature (solid circles) extracted from primary and secondary fragments are slightly different. The IMF multiplicity derivative is found at exactly the critical temperature for the primary fragments. But due to the secondary decay effect, the extracted critical point appears at slightly lower temperature for the secondary fragments. On the contrary, the measures of $M_2$ and the Fisher exponent $\tau$ predict an accurate critical temperature for the secondary fragments, but show slightly larger temperature for $M_2$ and lower temperature for $\tau$ for the primary fragments. The $\gamma_2$ is found to have similar accuracy in both the primary and secondary fragments, though a large error bar is obtained for primary fragments. In addition, the temperature error bars are also smaller for $M_2$, $\gamma_2$ and $\tau$ for the secondary fragments, which indicate these measures are more sensitive for the secondary fragments. The Zipf's law parameter $\xi$ shows a critical temperature slightly larger than that from the specific heat capacity for primary fragments. But due to its sharp response, the temperature error bar still does not cover the critical temperature from specific heat capacity. The result from secondary fragments is much worse for the Zipf's law parameter ($\xi$).

 From the above comparisons, we conclude that the total multiplicity derivative and NVZ are the best measures, which predict the critical point accurately with a minimal uncertainty both for the primary and secondary fragments.

\section*{V. Summary}

  The multiplicity derivatives, the moment parameters, the bimodal parameter, the fluctuation of maximum fragment charge number (NVZ), the Fisher exponent ($\tau$) and Zipf's law parameter ($\xi$) are examined as the measures to search for the liquid-gas phase transition in nuclear multifragmention processes within the framework of SMM. The sensitivities of these measures are studied. All these measures predict a critical signature at or near to the critical point extracted from the specific heat. Among these measures, the total multiplicity derivative and NVZ are found to be the best measures in accuracy and sensitivity for the first-order phase transition even after the secondary decay process. The IMF multiplicity derivative is found to be accurate in the primary fragments but show a slight deviation from the critical temperature in the secondary fragments. On the contrary, the $M_2$ and the Fisher exponent $\tau$ observables predict the critical point very well from the secondary fragments, but show a slight deviation for the primary fragments. The $\gamma_2$ shows similar accuracy (less than 0.1 MeV deviation) both for the primary and secondary fragments. The smaller temperature error bars for the secondary fragments indicate the measure of $M_2$, $\gamma_2$ and $\tau$ are more sensitive for the secondary fragments. A lower temperature is predicted by the bimodal parameter both for the primary and secondary fragments, while the Zipf's law parameter $\xi$ predicts higher temperatures both for the primary and secondary fragments. These investigations should provide a guide for future experiments and analyses in the study of nuclear liquid-gas phase transition.
  
\section*{Acknowledgments}
The authors thank A. S. Botvina for providing his code. This work is supported by the National Key Research and Development program (MOST 2016YFA0400501), the National Natural Science Foundation of China (Grant No. 11705242) and the Program for the CAS ``Light of West China'' (No. 29Y601030). This work is also supported by the US Department of Energy under Grant No. DE--FG02--93ER40773.

\end{document}